\pdfoutput=1
\documentclass[12pt,preprint]{aastex}
\usepackage[margin=0.75in]{geometry}
\usepackage{graphicx}
\usepackage{natbib}
\usepackage{aas_macros}
\usepackage[section] {placeins}
\usepackage{subfigure}
\usepackage{float}
\usepackage{color}

\def\msun{{\rm\,M_\odot}}

\def\msun{{\rm\,M_\odot}}

\def\h2{${\rm\,H_2}$}

\makeatletter 

\def\msun{{\rm\,M_\odot}}

\def\vol#1  {{{#1}{\rm,}\ }}

\newcount\refno
\newcount\rfno
\def\eq{$^{\the\refno\ }$\advance\refno by 1}
\def\ad{\advance\rfno by 1}

\def\clock{\count0=\time \divide\count0 by 60
     \count1=\count0 \multiply\count1 by -60 \advance\count1 by \time
     \number\count0:\ifnum\count1<10{0\number\count1}\else\number\count1\fi}

\def\myputfigure#1#2#3#4#5%
{\hskip0.03\textwidth\vskip#5pt
\makebox[0pt]{\hskip#2in
\includegraphics[width=#3\textwidth]{#1}}\vskip#4pt\hfill}

\newcount\refno
\newcount\rfno
\def\eq{$^{\the\refno\ }$\advance\refno by 1}
\def\ad{\advance\rfno by 1}

\definecolor{burntorange}{rgb}{1,0.4,0.2}

\begin{document}

\title{Quantifying Distributions of Lyman Continuum Escape Fraction}

\author{
Renyue Cen$^{1}$ and
Taysun Kimm$^{2}$
}

\footnotetext[1]{Princeton University Observatory, Princeton, NJ 08544;
 cen@astro.princeton.edu}

\footnotetext[2]{Princeton University Observatory, Princeton, NJ 08544;
kimm@astro.princeton.edu}

\begin{abstract} 
Simulations have indicated that most of the escaped Lyman continuum photons 
escape through a minority of solid angles with near complete transparency,
with the remaining majority of the solid angles largely opaque,
resulting in a very broad and skewed probability distribution function (PDF) of the 
escape fraction when viewed at different angles. Thus, the escape fraction of Lyman 
continuum photons of a galaxy observed along a line of sight merely represents 
the properties of the interstellar medium along that line of sight, which may be an 
ill-representation of true escape fraction of the galaxy averaged over its full sky.
Here we study how Lyman continuum photons escape from galaxies at $z=4-6$,
utilizing high-resolution large-scale cosmological radiation-hydrodynamic simulations.
We compute the PDF of 
the mean escape fraction ($\left<f_{\rm esc,1D}\right>$) averaged over 
mock observational samples, as a function of the sample size, compared 
to the true mean (had you an infinite sample size). We find that, when the sample 
size is small, the apparent mean skews to the low end. For example, for a true mean 
of 6.7\%, an observational sample of (2,10,50) galaxies at $z=4$ would have
have 2.5\% probability of obtaining the sample mean lower than 
$\left<f_{\rm esc,1D}\right>=$(0.007\%, 1.8\%, 4.1\%) 
and 2.5\% probability of obtaining the sample mean being greater than (43\%, 18\%, 11\%). 
Our simulations suggest that at least $\sim$ 100 galaxies should be stacked
in order to constrain the true escape fraction within 20\% uncertainty.
\end{abstract}

\section{Introduction}

A fraction of the Lyman continuum (LyC) photons generated by young massive stars is 
believed to escape from the host galaxies to enter the intergalactic space.
This is a fundamental quantity
to determine the epoch and pace of cosmological reionization,
provided that the universe is reionized by stars \citep[e.g.,][]{2000aGnedin,2003Cen}.
After the completion of cosmological reionization,
it plays another important role in determining the ultra-violet (UV) radiation background (on both sides of the Lyman limit)
in conjunction with another major source of UV photons - quasars - that progressively gains importance at lower redshift
\citep[e.g.,][]{2008FG,2014Fontanot}.

Observations of star-forming galaxies at high redshifts ($z\sim3$) suggest a wide range of the escape fraction 
of ionizing photons. While only a small fraction of LyC photons ($\lesssim$ a few percent) escapes 
from their host galaxies in the majority of the Lyman break galaxy samples, a non-negligible number 
of them ($\sim10\%$) shows high levels of LyC flux corresponding to  $\left<f_{\rm esc,1D}\right>\sim10\%$ 
\citep{shapley06,iwata09,nestor11,nestor13,2013Mostardi}. \citet{cooke14} claim that the mean escape fraction 
may be even higher ($\left<{\rm f_{esc},1D}\right>\sim 16\%$) if the observational sample is not biased 
toward the galaxies with a strong Lyman limit break.
It is not well understood quantitatively, however, what the probability distribution function (PDF) 
of the LyC escape fraction is and how a limited observational sample size with
individually measured escape fractions can be properly interpreted,
because of both possible large variations of the escape fraction from sightline to sightline for a given galaxy
and possible large variations from galaxy to galaxy.
The purpose of this {\it Letter}
is to quantify how LyC photons escape, in order to
provide a useful framework for interpreting and understanding
the true photon escape fraction given limited observational sample sizes.

\section{Simulations}\label{sec: sims}

To investigate how LyC photons escape from their host halos, we make use of the cosmological radiation 
hydrodynamic simulation performed using the Eulerian adaptive mesh refinement code, 
{\sc ramses} \citep[][ver. 3.07]{teyssier02,rosdahl13}.
The reader is referred to \citet[][, the FRU run]{2014Kimm} 
for details, where a detailed prescription for a new, greatly improved treatment of stellar feedback in the form 
of supernova explosion is given. Specifically, the new feedback model follows the dynamics of the explosion 
blast waves that capture the solution for all phases (from early free expansion to late snowplow), 
independent of simulation resolution and allow for anisotropic propagation.

The initial condition for the simulation is generated using the {\sc MUSIC} software \citep{2011Hahn},
with the WMAP7 parameters \citep{2011Komatsu}:
$(\Omega_{\rm m}, \Omega_{\Lambda}, \Omega_{\rm b}, h, \sigma_8, n_s  = \\
0.272, 0.728, 0.045, 0.702, 0.82, 0.96)$.
We adopt a large volume of $(25 {\rm Mpc/h})^3$ (comoving) to include the effect of large-scale 
tidal fields on the galaxy assembly. The entire box is covered with $256^3$ root grids, and 
high-resolution dark matter particles of mass $M_{\rm dm}=1.6\times10^5\,M_{\odot}$ are employed  
in the zoomed-in region of $3.8\times 4.8 \times 9.6$ Mpc$^3$. We allow for 12 more levels of grid 
refinement based on the density and mass enclosed within a cell in the zoomed-in region to have 
a maximum spatial resolution of 4.2 pc (physical). Star formation is modeled by creating normal 
and runaway particles in a dense cell ($n_H\ge 100\,{\rm cm^{-3}}$) with the convergent flow 
condition \citep[][the FRU run]{2014Kimm}. The minimum mass of a normal (runaway) star particle 
is $34.2\msun (14.6\msun)$. We use the mean frequency of Type II supernova explosions of 
$0.02 \msun^{-1}$, assuming the Chabrier initial mass function. Dark matter halos are identified 
using the HaloMaker \citep{2009Tweed}.

Eight consecutive snapshots are analyzed at each redshift ($3.96 \le z \le 4.00$, $4.92 \le z \le 5.12$, 
and $5.91 \le z \le 6.00$) to increase the sample size in our calculations.
At each snapshot there are $\approx$ 142, 137, and 104 halos in the halo mass 
range of $10^9 \le M_{\rm vir} < 10^{10}\,\msun$, and 15, 10, and 7 halos with mass 
$M_{\rm vir} \ge 10^{10}\msun$. The most massive galaxy at $z=4$ (5, 6) has stellar mass 
of $1.6\times 10^9\msun$ ($6.0\times 10^8$, $2.5\times 10^7\msun$), and host halo mass 
$8.8\times 10^{10}\msun$ ($5.2\times 10^{10}$, $4.1\times 10^{11}\,\msun$)

The escape fraction is computed as follows.
We cast $768$ rays per star particle and follow their propagation through the galaxy.
Each ray carries the spectral energy distribution (SED), including its LyC emission,
determined using {\sc Sturburst99~}\citep[][]{1999Leitherer}, given the age, metallicity, and mass of the star particle. 
The LyC photons are attenuated by neutral hydrogen \citep[][]{2006Osterbrock} and SMC-type dust 
\citep[][]{2007Draine} in the process of propagation. 
For a conservative estimate, we assume the dust-to-metal ratio of 0.4. 
We also simply assume that dust is destroyed in hot gas ($T>10^6~$K). 
We note that attenuation due to dust is only significant 
in the most massive galaxy ($M_{\rm star}=1.1\times 10^9\msun$, $\tau_d = 0.58$) in our sample. 
The second most massive galaxy ($M_{\rm star}=3.6\times 10^8\msun$) shows $\tau_d = 0.29$, 
meaning that it reduces the number of photons by only $<30\%$. 
Given that the dust-to-metal ratio is even smaller than 0.4 in low-metallicity systems 
\citep[][]{1998Lisenfeld, 2008Engelbracht, 2011Galametz, 2013Fisher}, 
it is likely that the attenuation by dust is even less significant in our simulated galaxies.
We define the true escape fraction of the galaxy as the ratio of the sum of all outward fluxes at the virial sphere 
to the sum of the initially emitted fluxes of all stellar particles in the galaxy;
we shall call this $f_{\rm esc,3D}$.
In addition, an observer at infinity at a random point in the sky of the galaxy
collects all LyC fluxes and defines the escape fraction along that particular line of sight;
this is called $f_{\rm esc,1D}$.

\section{Probability Distribution Functions of LyC Photon Escape Fraction}

\begin{figure}[!h]
\centering
\hskip -1.0cm
\resizebox{3.5in}{!}{\includegraphics[angle=0]{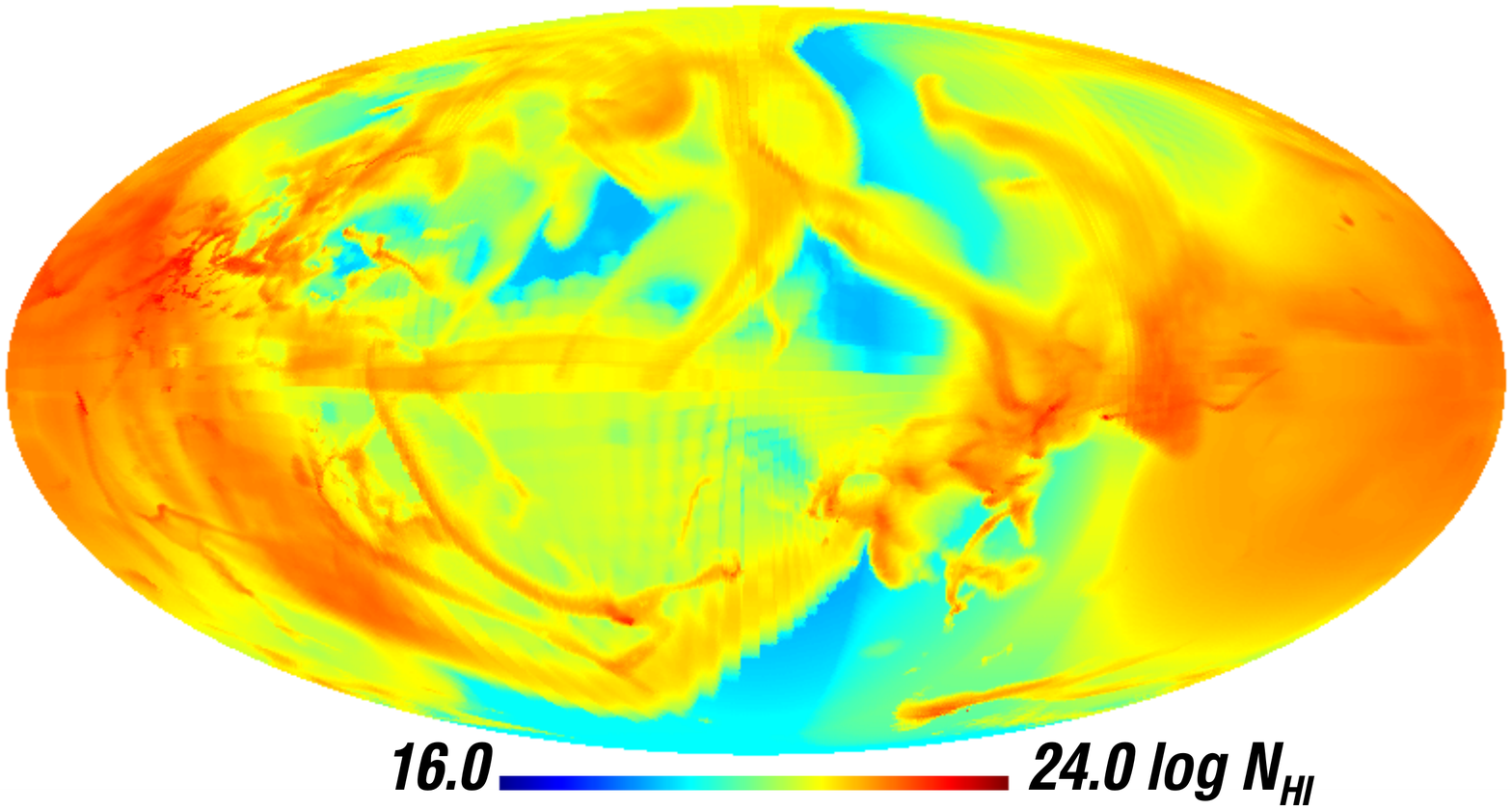}}
\resizebox{3.5in}{!}{\includegraphics[angle=0]{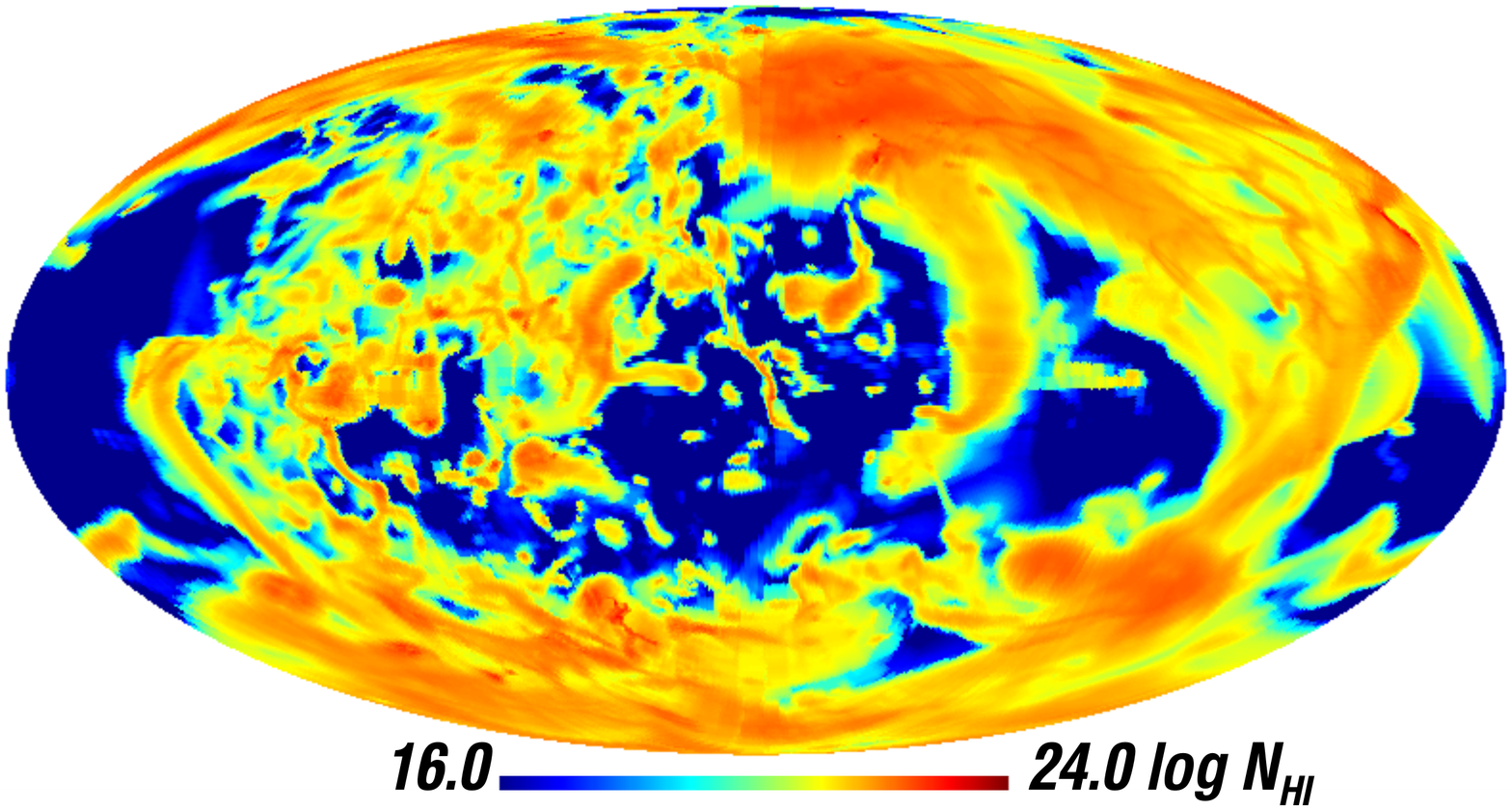}}
\vskip -1.5cm
\resizebox{3.5in}{!}{\includegraphics[angle=0]{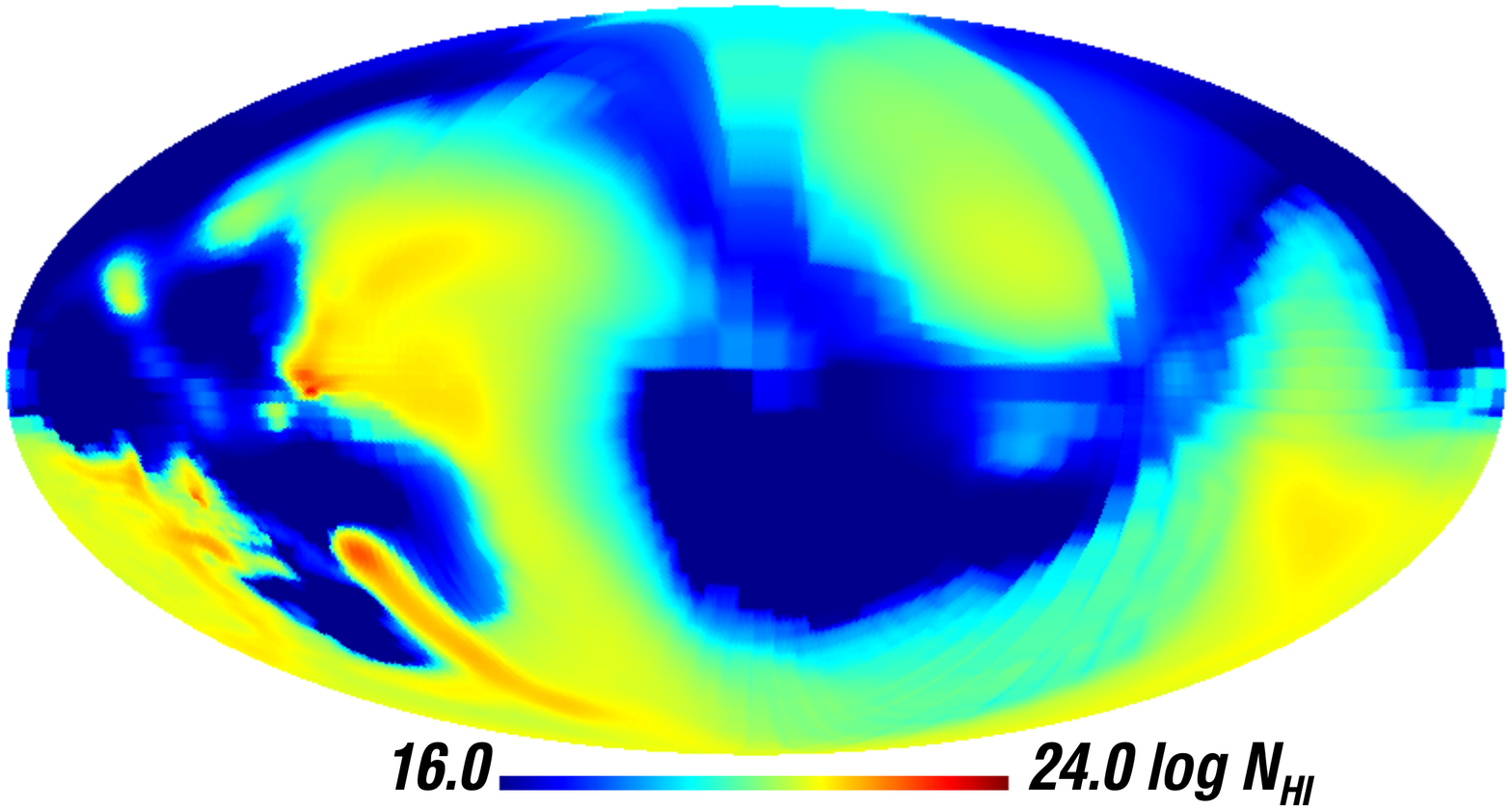}}
\vskip -1.0cm
\caption{
shows three examples of all-sky maps - the sky an observer sitting at the center of the galaxy would see -
of the neutral hydrogen for most massive ($M_{vir}=7.8\times 10^{10}\msun$, top left panel),
second massive ($6.1\times 10^{10}\msun$, top right panel),
and a smaller halo ($1.8\times 10^9\msun$, bottom).
The observer is placed at the center of the halo.
Note that the actual escape fraction presented later is computed
by ray-tracing LyC photons of all stellar particles spatially distributed
through the clumpy interstellar medium until escaping through the virial sphere.
The true escape fraction of LyC photons of these halos are 5.4\%, 12\%, and 5.0\%, respectively.
}
\label{fig:maps}
\end{figure}

It is useful to give a qualitative visual illustration of how LyC photons may escape from galaxies at $z=4$.
Figure~\ref{fig:maps}
shows three examples of an all-sky map - the sky an observer sitting at the center of the galaxy would see -
of the neutral hydrogen column density.
We note that 8 dex of dynamical range is plotted and recall that 
at the Lyman limit a neutral hydrogen column density of 
$\sim 10^{17}$cm$^{-2}$ would provide an optical depth of $\sim 1$.
As a result, LyC photons can only escape through 
highly ionized or evacuated ``holes" indicated by dark blue colors on the maps 
and the transition from near transparency to very opaque is fast.
This indicates that the escaping LyC photons are dominated by those that escape through 
completely unobscured channels and the amount of escaped LyC photons for a given galaxy 
depends strongly on the direction. 
Moreover, it is evident that, in addition to large variations from position to position on the sky for a given galaxy,
there are large variations of the overall column density structures from galaxy to galaxy.
For example, the galaxy in the top-left panel shows no transparent sky patches at all,
which is typical for galaxies during times of intense starburst as shown in \citet[Figure 4][]{2014Kimm}.
On the other hand, 
the galaxy in the bottom panel has large swaths of connected transparent patches that cover nearly one half 
of the sky, typical for galaxies 
at periods following the blowout of gas subsequent to intense starburst \citep[Figure 4][]{2014Kimm}.
This qualitative behavior is also found earlier in independent simulations by \citet[][]{2009Wise}.


\begin{figure}[!h]
\centering
\hskip -1.0cm
\resizebox{3.5in}{!}{\includegraphics[angle=0]{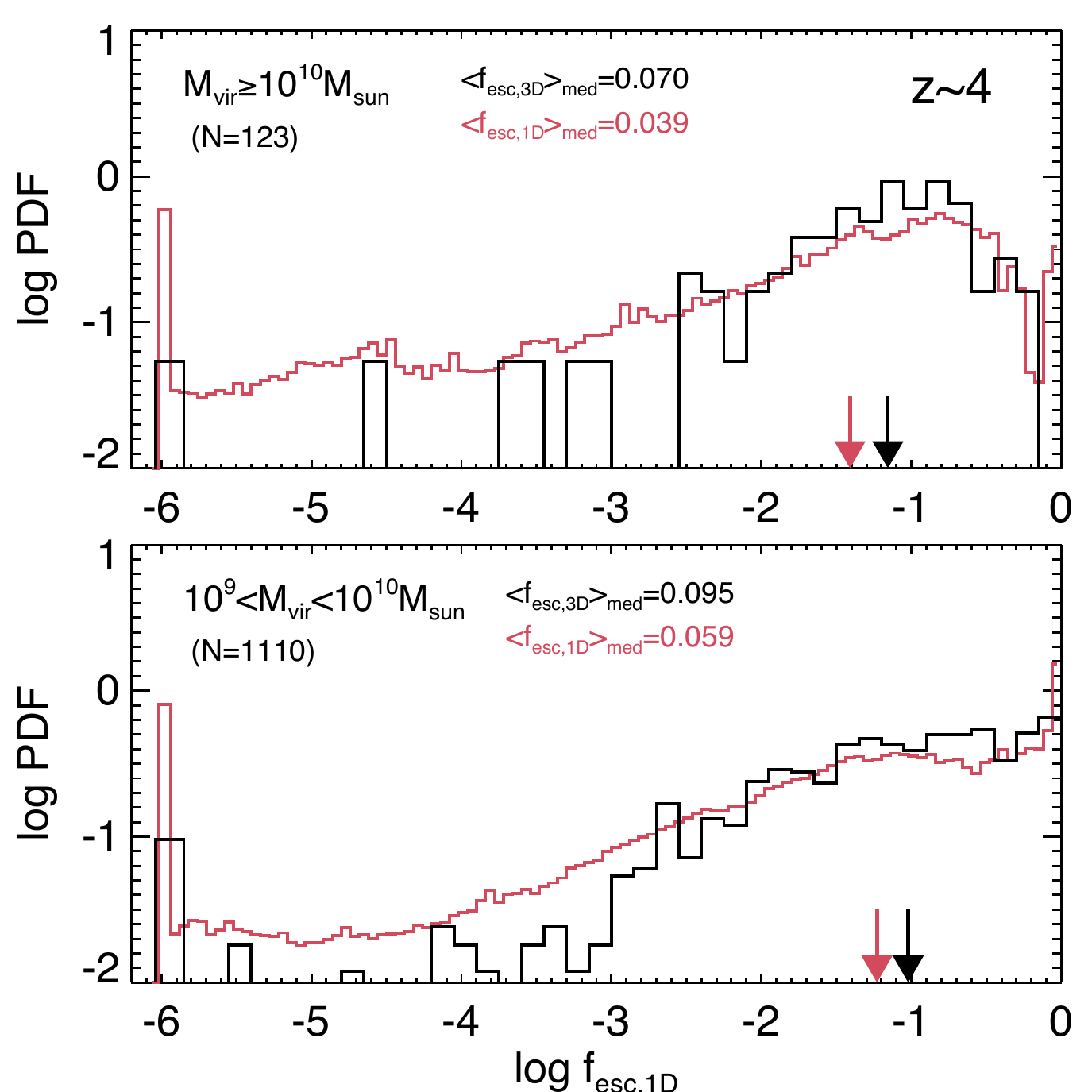}}
\hskip -0.0cm
\resizebox{3.5in}{!}{\includegraphics[angle=0]{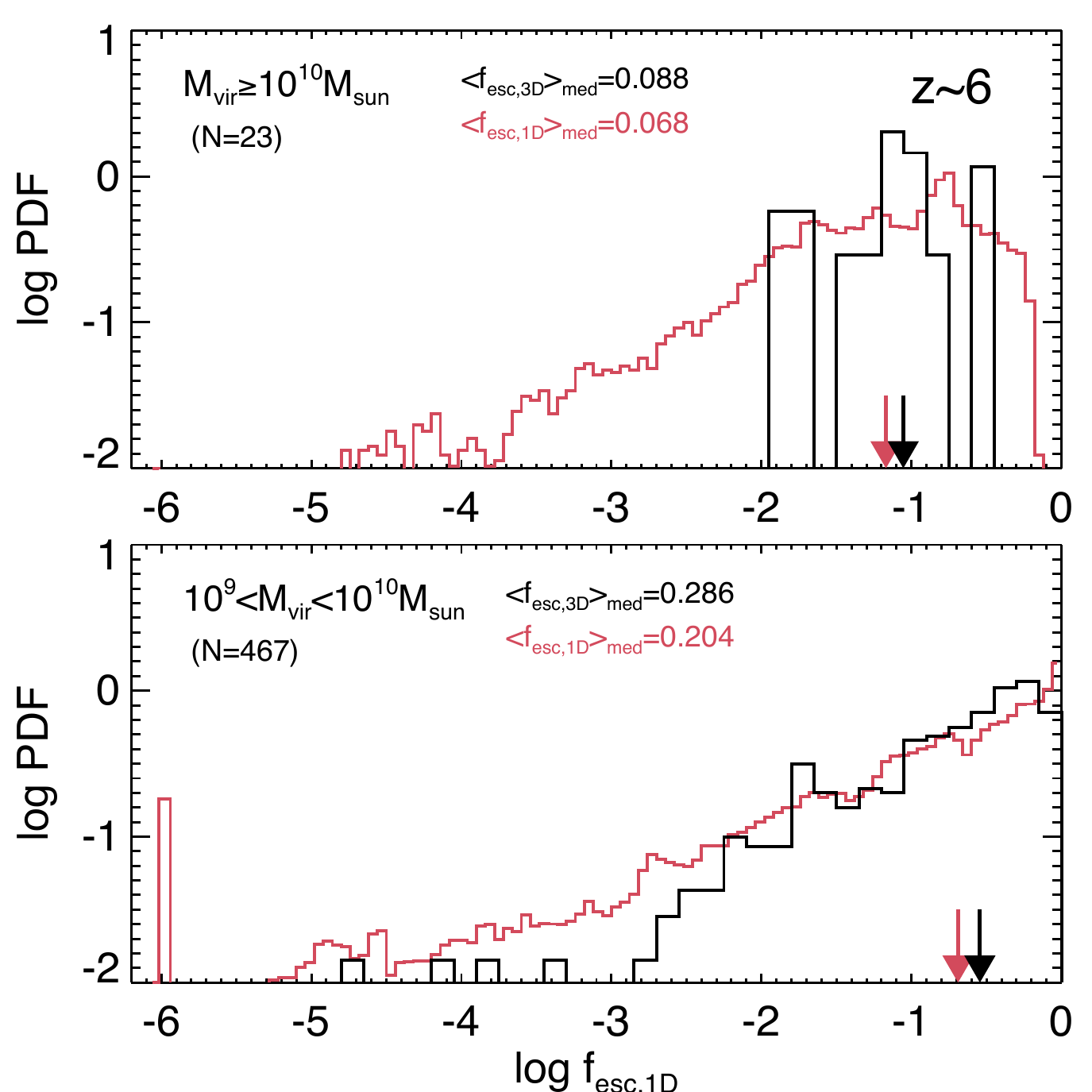}}
\vskip 0.5cm
\caption{
shows
the probability distribution of the apparent escape fraction for massive halos (top panel) and less massive halos (bottom panel)
for $z=4$ (left column) and $z=6$ (right column).
Black histograms show the distribution of the true escape fraction of each sample,
while red histograms show the PDF of the apparent escape fraction.
The median of the distribution is shown as arrows.
}
\label{fig:PDFfesc}
\end{figure}

Let us now turn to more quantitative results.
Figure~\ref{fig:PDFfesc} shows the probability distribution of the apparent escape fraction for massive 
halos (top) and less massive halos (bottom) at $z=4$ (left column) and $z=6$ (right column). 
Black histograms show the distribution of the true (3D) escape fraction of each sample (i.e., from the viewpoint of the overall intergalactic medium), 
while red histograms show the PDF of the apparent escape fraction (i.e., from the point of view of observers placed at a far distance). 
Note that the distribution of the true escape fraction is noisier than that of the apparent escape fraction 
due to the smaller sample size for the former, 
because for (3D) escape fraction each galaxy is counted once but for
the apparent escape fraction each galaxy is sampled many times.
In terms of the mean escape fraction, there is a trend that, at a given redshift,
the galaxies embedded in more massive halos tend to have a lower mean escape fraction.

There is also a weak trend that the escape fraction increases with redshift.
For example, the true (3D) median escape fractions are (7.0\%, 9.5\%) for the halos of masses 
($\ge 10^{10}, 10^9-10^{10})\msun$, respectively, at $z=4$; 
the true (3D) median escape fractions are (8.8\%, 29\%) for the halos of masses 
($\ge 10^{10}, 10^9-10^{10})\msun$, respectively, at $z=6$. 
Upon a close examination we suggest that the redshift dependence can be attributed, in part, to the following findings.
At a given halo mass, the specific star formation rate decreases 
with decreasing redshifts at $4\le z \le 6$. 
As star formation becomes less episodic at lower redshifts, it takes longer to blow out the star-forming clouds via SNe.
Consequently, a larger fraction of LyC photons is absorbed by their birth clouds. We also find that the specific star 
formation rate does not change notably at $z>6$ while the mean density of the halo increases with redshifts,
explaining an opposite trend found in \citet[][]{2014Kimm} at $7\le z \le 11$. 
The predicted median 
escape fraction for halos with $10^9\le M_{\rm halo} \le 10^{11}\,M_{\odot}$ at $4\le z \le 6$ is generally smaller than the previous studies 
\citep[$10-30\%$][]{2010Razoumov, 2014Yajima}, although the trend at $z<6$ is broadly consistent with 
\citet[][]{2010Razoumov}.
Nevertheless, it is prudent to bear in mind that
the sometimes conflicting results and trends among studies with respect to redshift 
may be in part due to still limited galaxy sample sizes.

\begin{figure}[!h]
\centering
\hskip -1.0cm
\resizebox{3.5in}{!}{\includegraphics[angle=0]{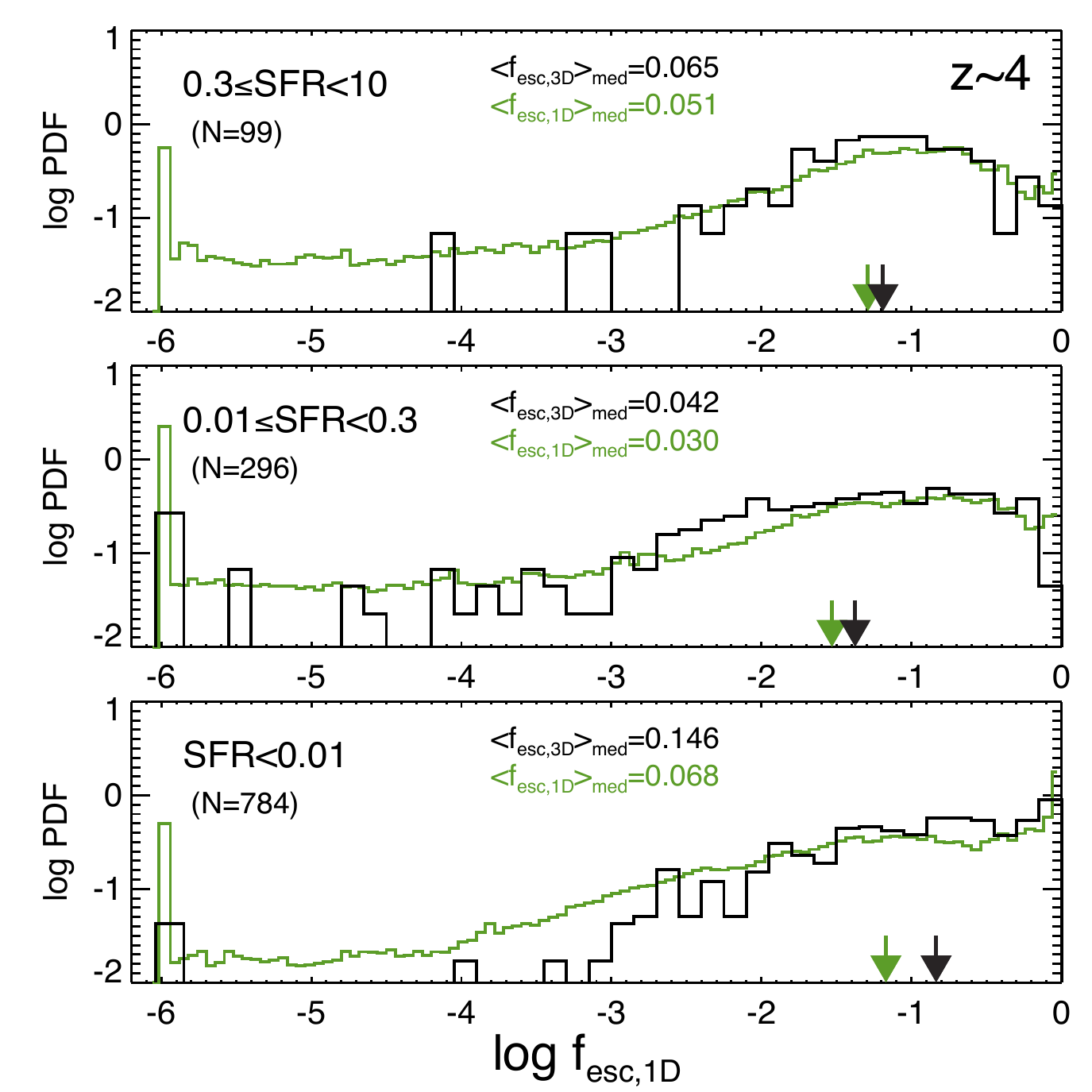}}
\hskip -0.0cm
\resizebox{3.5in}{!}{\includegraphics[angle=0]{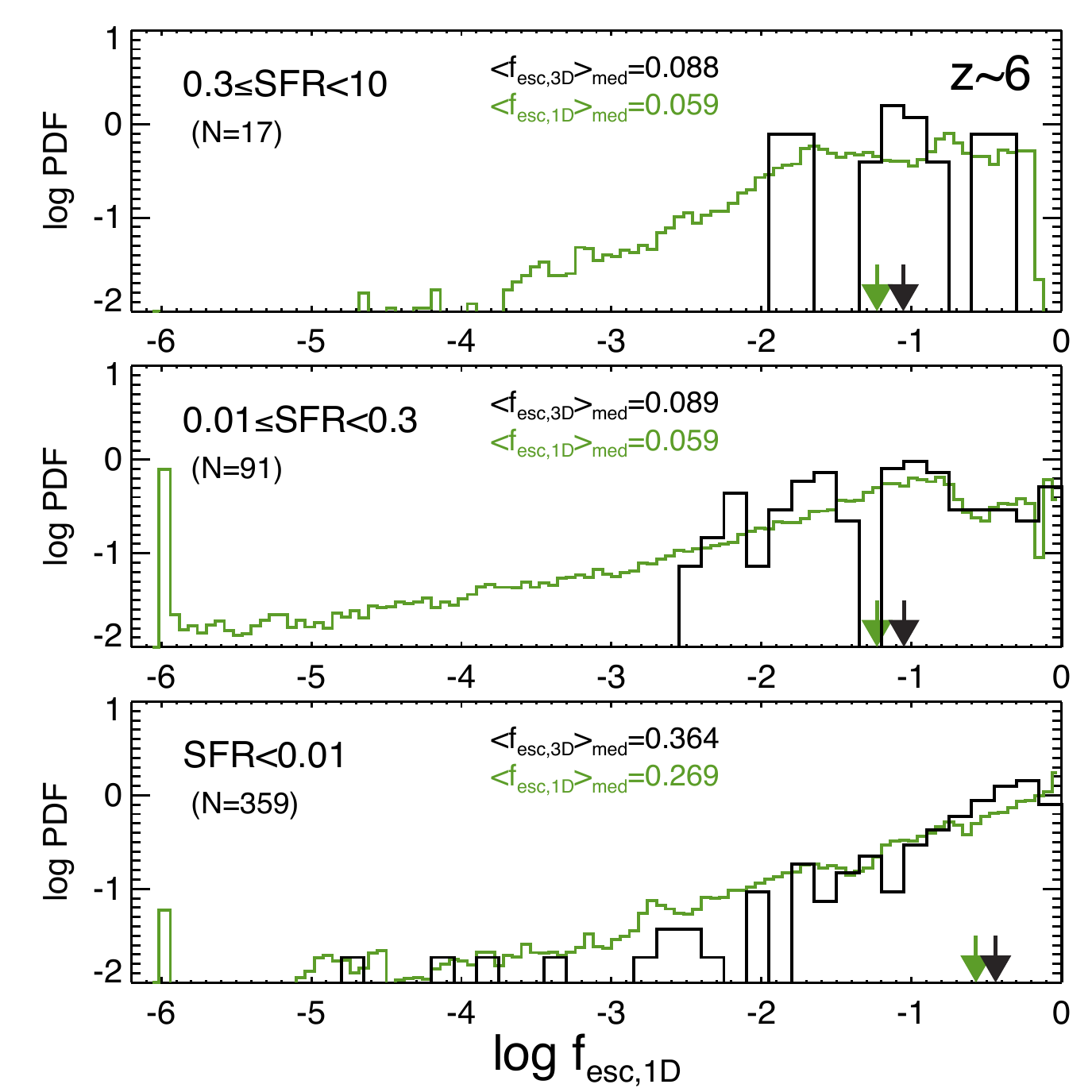}}
\vskip 0.5cm
\caption{
is similar to Figure~\ref{fig:PDFfesc}, except the galaxy sample is subdivided according to their star formation rates,
SFR=$0.3$ to $10\msun$/yr (top panel),
$0.01$ to $0.3\msun$/yr (middle panel),
and $<0.01\msun$/yr (bottom panel).
}
\label{fig:PDFfesc2}
\end{figure}

Figure~\ref{fig:PDFfesc2} is similar to Figure~\ref{fig:PDFfesc}, except the galaxy sample is 
subdivided according to their star formation rates (SFRs). We note that this division according to SFRs
introduces subtle degeneracies.
For example, a lower SFR does not necessarily correspond to a less massive galaxy;
instead, a lower SFR may correspond to the phase of a galaxy between two star formation bursts.
The significantly higher escape fraction for the lowest SFR bin (bottom panels)
is, to the most part, due to a post-starburst phase when the interstellar medium has been
cleared out by the preceding burst and SFR has abated, as noted in \citet[][]{2014Kimm}.
Thus, if we do not consider the lowest SFR bin, it seems that the mean escape fraction 
does not strongly depend on SFR at $z=4-6$.
The escape fraction from the most actively star-forming galaxy sample with $0.3 < {\rm SFR} < 10\msun$/yr  
be compared with that of Lyman alpha emitters or faint LBGs. Our simulations suggest that the median 
$f_{\rm esc,1D}$ of the sample is 5.1\%, which is consistent with $f_{\rm esc,1D}=5-15\%$ inferred 
from narrow band filter imaging observations of 91 LAEs  (Mostardi et al. 2013). We note that,  
given the wide distribution of the simulated apparent escape fraction ($0.1\%\lesssim f_{\rm esc,1D} 
\lesssim 23\%$ or $0.001\%\lesssim f_{\rm esc,1D} \lesssim 47\%$ for the 1 and 1.5 $\sigma$ range, respectively),
our results are also compatible with the individual detection of LyC fluxes from 7 LBGs 
(Iwata et al. 2009, $5.5\lesssim f_{\rm esc,1D} \lesssim 55\%$ for the intrinsic UV to LyC flux ratio of 3.).

\begin{figure}[!h]
\centering
\vskip 0.5cm
\hskip -1.0cm
\resizebox{3.5in}{!}{\includegraphics[angle=0]{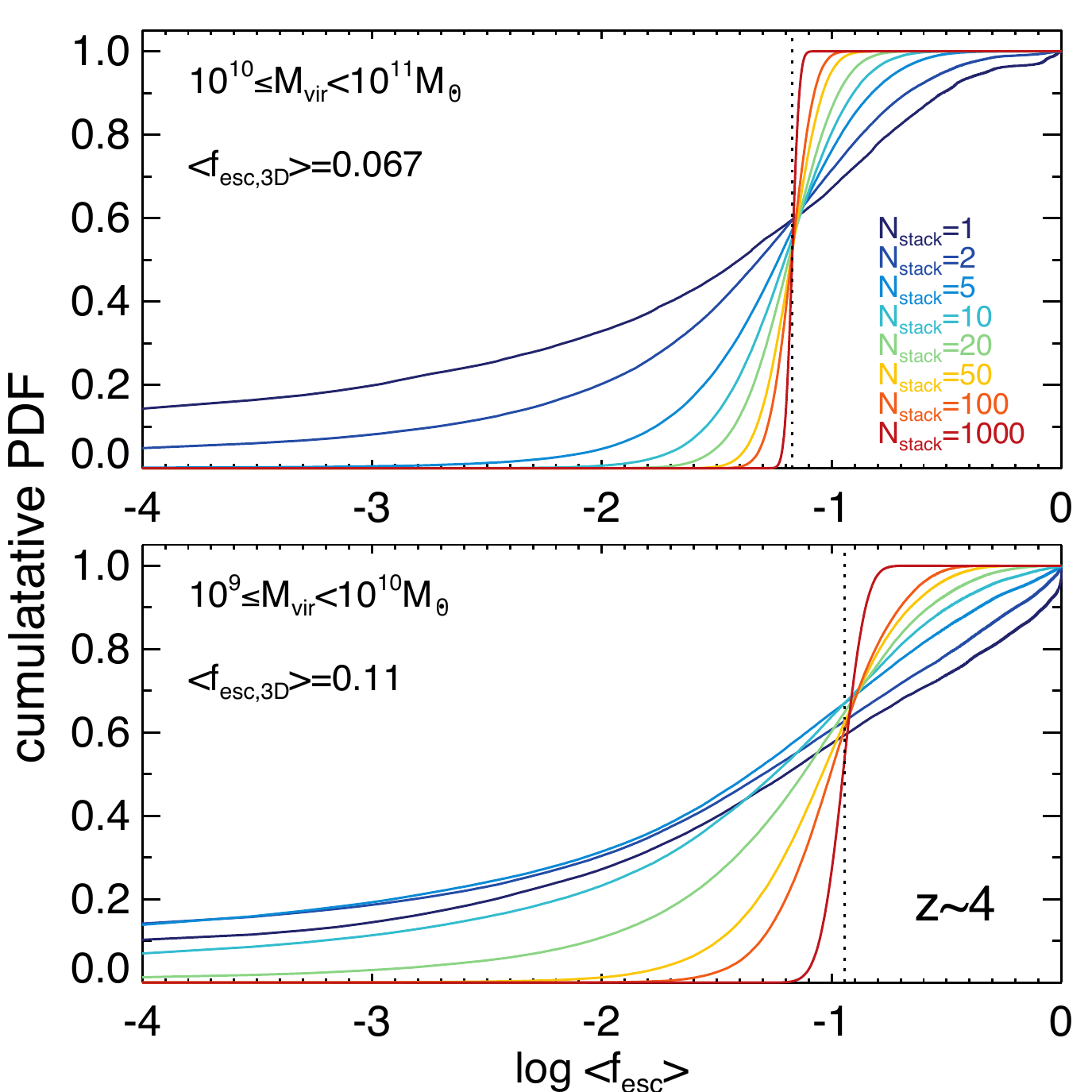}}
\hskip 0.0cm
\resizebox{3.5in}{!}{\includegraphics[angle=0]{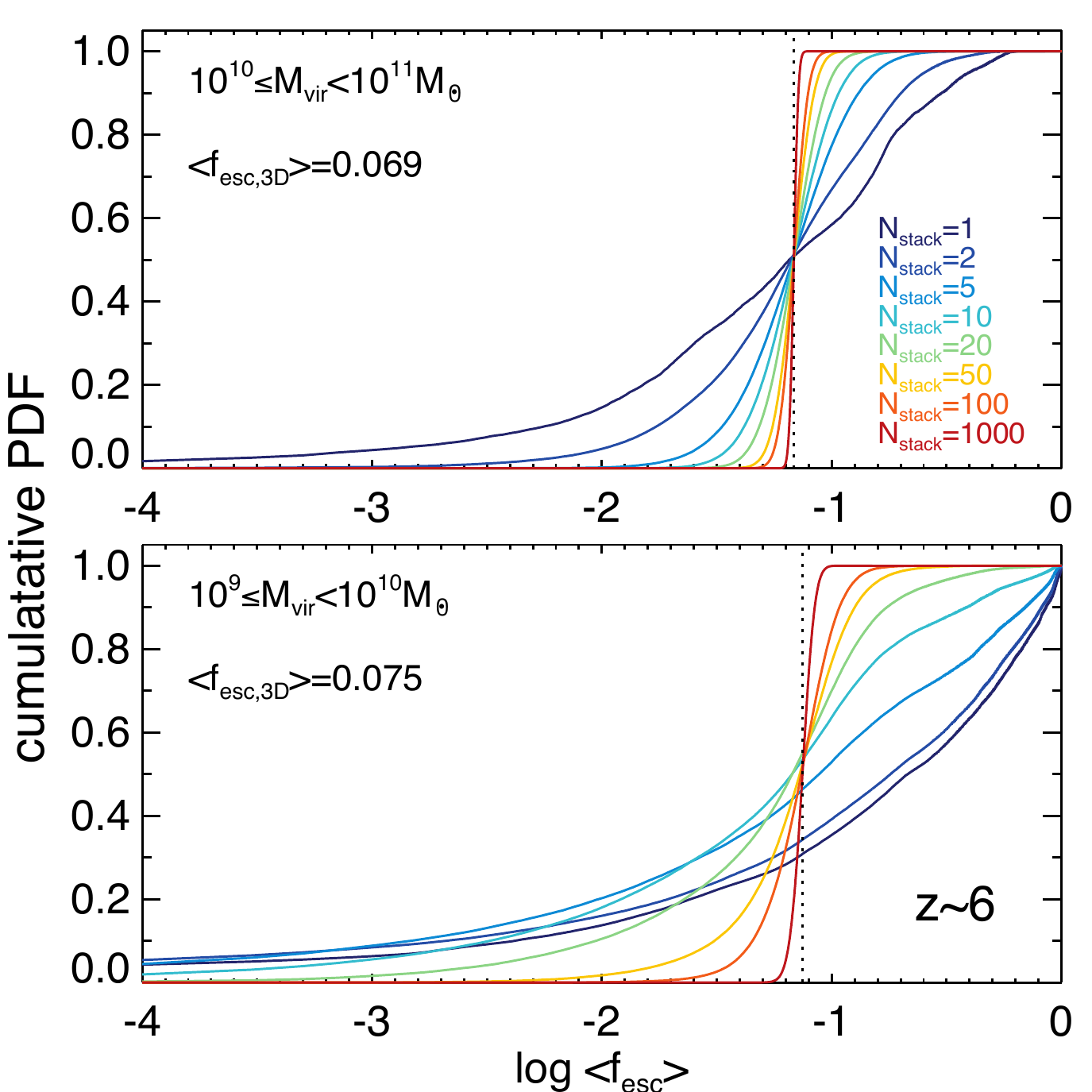}}
\vskip 2.5cm
\hskip -1.0cm
\resizebox{3.5in}{!}{\includegraphics[angle=0]{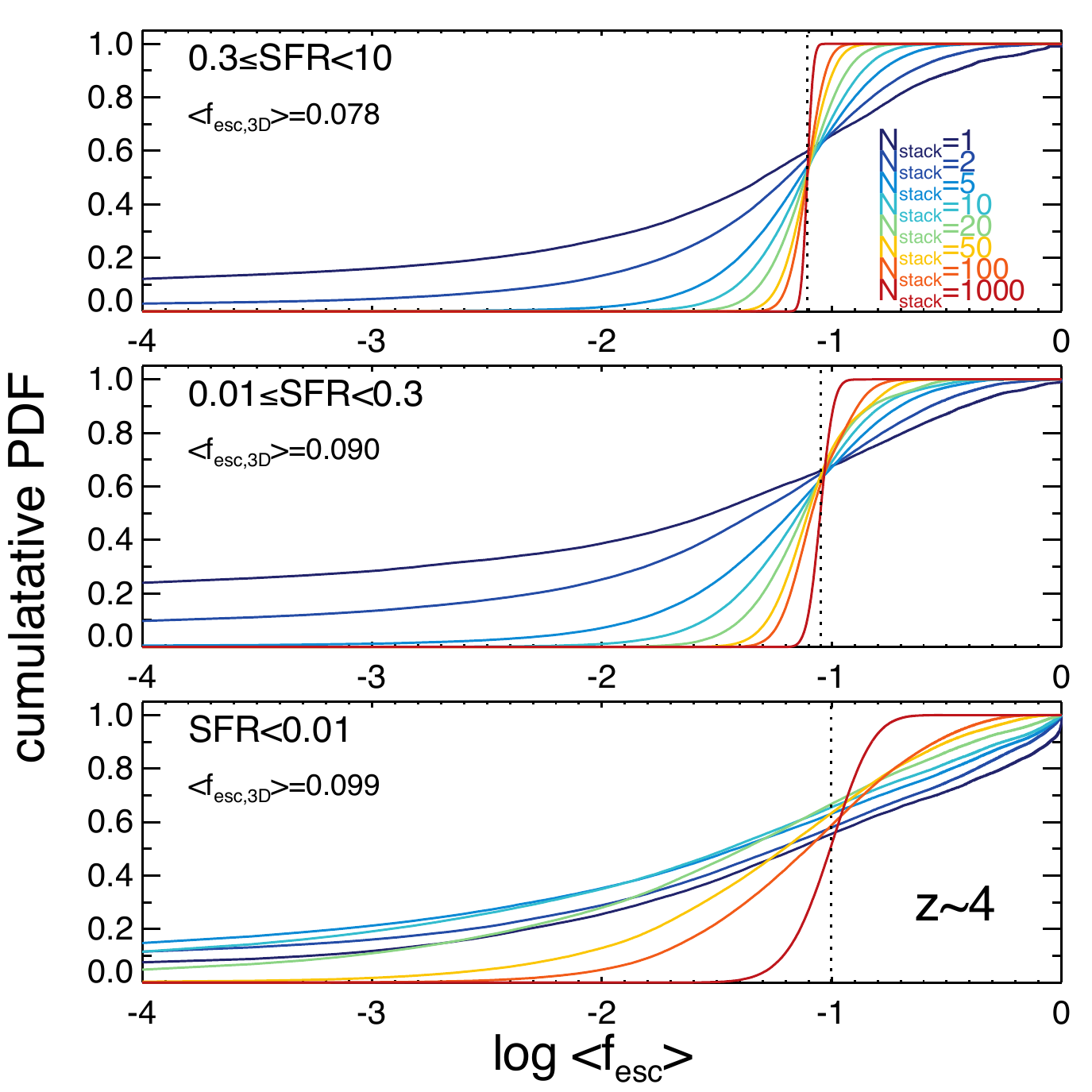}}
\hskip 0.0cm
\resizebox{3.5in}{!}{\includegraphics[angle=0]{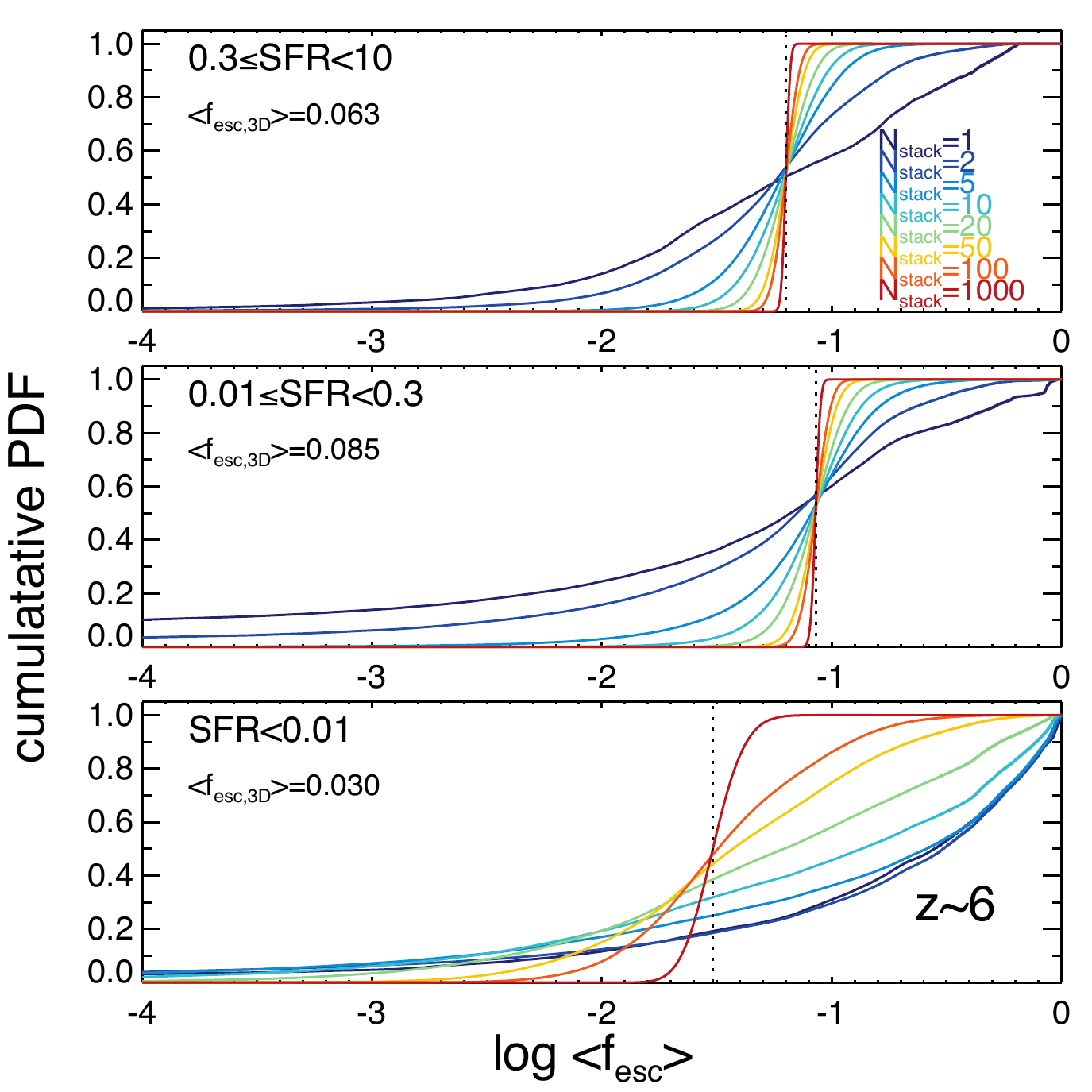}}
\vskip -0.0cm
\caption{
{\color{burntorange}\bf Top two panels}
show the probability distribution function of the apparent mean for a given observational sample size $N_{\rm stack}$
for the high mass (top) and low mass (bottom) sample, respectively.
The mean is computed by weighting
the number of photons produced in each galaxies to mimic the stacking of the SED in observations.
The true mean of the distribution is denoted in each panel.
{\color{burntorange}\bf Bottom two panels}
are the same as the top two panels, but for the subsamples with different star formation rates, as indicated in the legend.
}
\label{fig:prob}
\end{figure}

Evidently, the distribution of the apparent LyC escape fraction is very broad and skewed toward the lower end.
The reason for this behavior is understandable. 
In the case of the galaxies with low ${\rm f_{esc,3d}}$ values, 
the LyC photons escape normally through transparent holes with small solid angles.
Since not all of these holes are seen to an observer, the distribution of ${\rm f_{esc,1d}}$ for individual galaxies 
tends to get skewed toward the lower end of the distribution. 
As a result, the medians of the two distributions, shown as arrows in Figure~\ref{fig:PDFfesc}, 
are about a factor of $\sim2$ smaller than the mean.
More importantly, it suggests that an observational sample of limited size may 
underestimate the true mean escape fraction.
The top two panels of Figure~\ref{fig:prob}
show the probability distribution function of the apparent mean for a given observational sample size $N_{\rm stack}$
for the high mass (top) and low mass (bottom) sample, respectively.
We compute the apparent mean of a sample of galaxies using LyC photon (or SFR)-weighted mean
escape fraction, which is exactly equivalent to stacking the galaxies.
The bottom two panels of Figure~\ref{fig:prob}
are similar to top two panels in Figure~\ref{fig:prob}, for the subsamples with different star formation rates.
What we see in these figures is that the probability distribution is rather broad.
It is thus clear that it is not a robust exercise to try to infer
the mean escape fraction based on a small sample ($\le 10$) of galaxies,
whether individually measured or through stacking.

Table 1 provides a quantitative assessment of the uncertainties,
which shows the $1$ and $2\sigma$ probability intervals of fractional lower and upper deviations 
from the true mean escape fraction.
Some relatively mild trends are seen that are consistent with earlier observations of the figures.
Specifically, the convergence to the true mean escape fraction in terms of sample sizes 
is faster towards high redshift, towards higher halo mass, and towards higher star formation rates.
Let us take a few numerical examples.
We see that with a sample of 50 galaxies of halo mass in the range of $(10^{10}-10^{11})\msun$ at $z=4$ 
the $2\sigma$ fractional range of the escape fraction is 58\% to 159\%,
which improves to a range of 68\% to 140\% when a sample of 100 galaxies is used.
Note that the observations of \citet[][]{2013Mostardi} 
have 49 Lyman break galaxies and 91 Lyman alpha emitters at $z\sim 2.85$.
At $z=6$ for the $(10^{10}-10^{11})\msun$ halo mass range, 
we see that with a sample of 20 galaxies, 
the $2\sigma$ fractional range of the escape fraction is 59\% to 161\%,
comparable to that of a sample of 50 galaxies at $z=4$, as a result of benefiting from
the faster convergence at higher redshift.
On the other hand, at $z=5$ for the $(0.3-10)\msun$yr$^{-1}$ star formation rate range, 
the $2\sigma$ fractional range of the escape fraction is 56\% to 163\% with a sample of 20 galaxies,
which is improved to 71\% to 137\% with a sample of 50 galaxies.

Finally, we note that the actual observed Lyman continuum escape fraction has additionally 
suffered from possible absorbers in the intergalactic medium, primarily Lyman limit systems.
Since the background galaxy and the foreground absorbers are physically unrelated,
we may consider the effects from the internal factors in galaxies and those from the intergalactic medium
completely independent.
Thus, in this case, assuming no knowledge of the foreground absorbers,
the overall distribution would be the convolution of the two, resulting a still broader overall distribution
than derived above considering internal factors alone.
In reality, however, one may be able to remove, to a large degree, the Lyman continuum opacity due to
intergalactic absorbers by making use of a tight correlation between Ly$\alpha$ and LyC absorption
\citep[][]{2008Inoue}.

\section{Conclusions}

We have simulated a significant sample of galaxies that are resolved at 4 parsec scales,
important for capturing the structure of the interstellar medium \citep[e.g.,][]{2006Joung}.
We have also implemented a much improved supernova feedback method that 
captures all phases of the Sedov-Taylor explosion solution and has been shown to 
yield the correct final momentum driven by the explosion regardless of the numerical resolution \citep[][]{2014Kimm}.
An adequate treatment of both these two requirements is
imperative, before one can start properly addressing the issue of LyC escape,
because most of the escape LyC photons escape through ``holes" in the interstellar medium, instead of
them uniformly leaking out in a ``translucent" medium.
In \citet[][]{2014Kimm} we address the escape fraction for galaxies at the epoch of reionization,
to provide the physical basis for stellar reionization.

Here we quantify the distribution of escape fraction for galaxies as a whole,
at a range of redshift from $z=4$ to $z=6$.
In general, it is found that
the LyC escape fraction depends strongly 
on the view angle of the observer and the overall distribution of the escape fraction sampled over
many sightlines is very broad.
The distribution narrows with increasing halo mass or SFR or redshift.
This broad distribution introduces large sampling uncertainties, when the galaxy sample size is limited.
For example, a sample of 50 galaxies of halo mass in the range of $(10^{10}-10^{11})\msun$ at $z=4$ 
produces the $2\sigma$ fractional range of the escape fraction of 58\%-159\%.
At $z=5$ a sample of 20 galaxies with star formation rate in the range of $(0.3-10)\msun$yr$^{-1}$ 
gives the $2\sigma$ fractional range of the escape fraction is 56\%-163\%.
Our analysis suggests that at least on order of tens of galaxies 
is needed, before one is confident at the $2\sigma$ level that the mean escape fraction measured 
does not deviate from the truth by 30-50\% at $z=4-6$ for galaxies hosted by halos of mass
in the range $10^{10}-10^{11}\msun$.

{\tiny
\begin{table*}
\centering
\vskip -1.8cm
\caption{1 and $2\sigma$ ranges of LyC escape fraction in terms of sample size, stellar mass and redshift}
\begin{tabular}{cccc}
\hline
\multicolumn{1}{c}{} & \multicolumn{1}{c}{$z=4$} & \multicolumn{1}{c}{$z=5$} & \multicolumn{1}{c}{$z=6$} \\
$N_{stack}$ & ($f_{-2\sigma},f_{-1\sigma}$,$f_{+1\sigma},f_{+2\sigma}$)/$\left<f_{\rm esc}\right>$& ($f_{-2\sigma},f_{-1\sigma}$,$f_{+1\sigma},f_{+2\sigma}$)/$\left<f_{\rm esc}\right>$& ($f_{-2\sigma},f_{-1\sigma}$,$f_{+1\sigma},f_{+2\sigma}$)/$\left<f_{\rm esc}\right>$\\
\hline
\multicolumn{4}{c}{ $10^{10} \leq M_{\rm vir} / M_{\odot} < 10^{11}$ }\\
1 & (0.001, 0.004,  3.21,  12.4) & (0.002, 0.097, 4.57,  14.6) & (0.004, 0.162, 3.19,  7.16) \\
2 & (0.001, 0.091, 2.34,  6.14) & (0.011, 0.231,  2.56,  7.45) & (0.092, 0.358,  2.41,  4.58) \\
5 & (0.091, 0.352,  1.81,  3.48) & (0.151, 0.470,  1.87,  3.36) & (0.311, 0.577,  1.65,  2.63) \\
10 & (0.251, 0.509,  1.59,  2.56) & (0.317, 0.601,  1.59,  2.48) & (0.460, 0.682,  1.42,  1.96) \\
20 & (0.405, 0.634,  1.43,  2.02) & (0.470, 0.703,  1.39,  1.92) & (0.588, 0.763,  1.28,  1.61) \\
50 & (0.580, 0.755,  1.27,  1.59) & (0.642, 0.805, 1.24,  1.51) & (0.717, 0.848,  1.17,  1.36) \\
100 & (0.684, 0.823,  1.18,  1.40) & (0.738, 0.859,  1.16,  1.34) & (0.795, 0.889,  1.12,  1.24) \\
1000 & (0.892, 0.942,  1.05,  1.11) & (0.910, 0.951,  1.05,  1.10) & (0.930, 0.965,  1.03,  1.07) \\
\hline
\multicolumn{4}{c}{ $10^{9} \leq M_{\rm vir} / M_{\odot} < 10^{10}$ }\\
1 & (0.001, 0.013,  5.37,  8.79) & (0.001, 0.264,  11.0,  13.4) & (0.001, 0.194,  9.98,  13.1) \\
2 & (0.001, 0.003,  3.86,  8.26) & (0.001, 0.155,  10.8,  13.3) & (0.001, 0.132,  9.31,  12.7) \\
5 & (0.001, 0.003,  2.58,  7.38) & (0.001, 0.061,  9.42,  13.0) & (0.001, 0.074,  6.97,  12.3) \\
10 & (0.001, 0.029, 2.03,  5.68) & (0.001, 0.080,  4.87,  12.3) & (0.001, 0.108,  3.05,  10.4) \\
20 & (0.006, 0.151, 1.79,  4.03) & (0.012, 0.185,  2.44,  7.94) & (0.023, 0.231,  1.85,  4.86) \\
50 & (0.131, 0.369, 1.58,  2.97) & (0.134, 0.445,  1.78,  3.23) & (0.170, 0.536,  1.49,  2.33) \\
100 & (0.277, 0.510, 1.49,  2.41) & (0.320, 0.601, 1.52,  2.29) & (0.419, 0.686,  1.33,  1.77) \\
1000 & (0.686, 0.820, 1.18,  1.38) & (0.745, 0.866, 1.14,  1.30) & (0.812, 0.900, 1.10,  1.20) \\
\hline
\multicolumn{4}{c}{ $0.3 \leq {\rm SFR} < 10$ }\\
1 & (0.001, 0.014,  2.89,  9.80) & (0.002, 0.106,  3.93,  14.3) & (0.008, 0.185, 4.74,  9.28) \\
2 & (0.001, 0.170,  2.19,  5.18) & (0.005, 0.272,  2.15,  5.53) & (0.062, 0.305, 2.28,  5.49) \\
5 & (0.165, 0.443,  1.76,  3.07) & (0.216, 0.538,  1.62,  2.64) & (0.277, 0.555, 1.58,  2.43) \\
10 & (0.330, 0.583, 1.54,  2.31) & (0.410, 0.666, 1.42,  1.99) & (0.447, 0.672, 1.39,  1.89) \\
20 & (0.476, 0.690, 1.36,  1.84) & (0.562, 0.760, 1.29,  1.63) & (0.583, 0.764, 1.26,  1.58) \\
50 & (0.636, 0.797, 1.23,  1.49) & (0.708, 0.843, 1.17,  1.37) & (0.719, 0.846, 1.16,  1.34) \\
100 & (0.731, 0.853, 1.16,  1.33) & (0.787, 0.888, 1.12,  1.25) & (0.795, 0.890, 1.11,  1.23) \\
1000 & (0.907, 0.953, 1.05,  1.10) & (0.929, 0.964, 1.04,  1.07) & (0.931, 0.963, 1.03,  1.07) \\
\hline
\multicolumn{4}{c}{ $0.01 \leq {\rm SFR} < 0.3$ }\\
1 & (0.001, 0.001,  2.87,  8.75) & (0.001, 0.009,  3.74,  12.1) & (0.001, 0.025,  3.99,  10.2) \\
2 & (0.001, 0.023,  2.21,  5.40) & (0.001, 0.108,  2.43,  6.76) & (0.001, 0.121,  2.02,  5.28) \\
5 & (0.034, 0.237,  1.75,  3.70) & (0.098, 0.388,  1.83,  3.23) & (0.100, 0.450,  1.63,  2.87) \\
10 & (0.161, 0.395, 1.52,  3.12) & (0.271, 0.551,  1.58,  2.39) & (0.325, 0.623,  1.42,  2.06) \\
20 & (0.304, 0.515, 1.38,  2.79) & (0.434, 0.669,  1.40,  1.91) & (0.513, 0.731,  1.29,  1.65) \\
50 & (0.468, 0.638, 1.35,  2.10) & (0.605, 0.781,  1.24,  1.53) & (0.684, 0.830,  1.18,  1.38) \\
100 & (0.569, 0.711, 1.32,  1.75) & (0.709, 0.842, 1.17,  1.36) & (0.769, 0.877,  1.13,  1.26) \\
1000 & (0.813, 0.896, 1.10,  1.22) & (0.900, 0.946, 1.05,  1.11) & (0.924, 0.961, 1.04,  1.08) \\
\hline
\multicolumn{4}{c}{ ${\rm SFR} < 0.01$ }\\
1 & (0.001, 0.025, 6.95,  10.1) & (0.001, 0.330, 9.99,  12.0) & (0.003, 0.654, 25.5,  32.3) \\
2 & (0.001, 0.009, 5.31,  9.64) & (0.001, 0.263, 10.1,  11.8) & (0.003, 0.674, 25.5,  31.5) \\
5 & (0.001, 0.001, 4.33,  9.36) & (0.001, 0.097, 9.73,  11.7) & (0.003, 0.273, 24.9,  31.1) \\
10 & (0.001, 0.005, 3.66,  9.15) & (0.001, 0.083, 8.51,  11.4) & (0.003, 0.220, 21.3,  30.1) \\
20 & (0.001, 0.025, 2.78,  8.18) & (0.011, 0.124, 4.98,  10.5) & (0.023, 0.246, 14.0,  27.3) \\
50 & (0.015, 0.132, 2.31,  6.04) & (0.054, 0.261, 2.37,  6.66) & (0.095, 0.345, 5.02,  14.8) \\
100 & (0.063, 0.239, 2.20,  4.65) & (0.133, 0.393, 1.96,  4.43) & (0.200, 0.463, 3.00,  7.21) \\
1000 & (0.468, 0.692, 1.38,  1.84) & (0.585, 0.765, 1.28,  1.64) & (0.614, 0.779, 1.30,  1.70) \\
\hline
\end{tabular}
\end{table*}
}

\vskip 1cm

We thank Rogier Windhorst for useful discussion.
Computing resources were in part provided by the NASA High-
End Computing (HEC) Program through the NASA Advanced
Supercomputing (NAS) Division at Ames Research Center.
The research is supported in part by NSF grant AST-1108700 and NASA grant NNX12AF91G.


\begin{thebibliography}{28}
\expandafter\ifx\csname natexlab\endcsname\relax\def\natexlab#1{#1}\fi

\bibitem[{{Cen}(2003)}]{2003Cen}
{Cen}, R. 2003, \apj, 591, 12

\bibitem[{{Cooke} {et~al.}(2014){Cooke}, {Ryan-Weber}, {Garel}, \& {Gonzalo
  Diaz}}]{cooke14}
{Cooke}, J., {Ryan-Weber}, E.~V., {Garel}, T., \& {Gonzalo Diaz}, C. 2014,
  ArXiv e-prints

\bibitem[{{Draine} {et~al.}(2007){Draine}, {Dale}, {Bendo}, {Gordon}, {Smith},
  {Armus}, {Engelbracht}, {Helou}, {Kennicutt}, {Li}, {Roussel}, {Walter},
  {Calzetti}, {Moustakas}, {Murphy}, {Rieke}, {Bot}, {Hollenbach}, {Sheth}, \&
  {Teplitz}}]{2007Draine}
{Draine}, B.~T., {Dale}, D.~A., {Bendo}, G., {Gordon}, K.~D., {Smith},
  J.~D.~T., {Armus}, L., {Engelbracht}, C.~W., {Helou}, G., {Kennicutt}, Jr.,
  R.~C., {Li}, A., {Roussel}, H., {Walter}, F., {Calzetti}, D., {Moustakas},
  J., {Murphy}, E.~J., {Rieke}, G.~H., {Bot}, C., {Hollenbach}, D.~J., {Sheth},
  K., \& {Teplitz}, H.~I. 2007, \apj, 663, 866

\bibitem[{{Engelbracht} {et~al.}(2008){Engelbracht}, {Rieke}, {Gordon},
  {Smith}, {Werner}, {Moustakas}, {Willmer}, \& {Vanzi}}]{2008Engelbracht}
{Engelbracht}, C.~W., {Rieke}, G.~H., {Gordon}, K.~D., {Smith}, J.-D.~T.,
  {Werner}, M.~W., {Moustakas}, J., {Willmer}, C.~N.~A., \& {Vanzi}, L. 2008,
  \apj, 678, 804

\bibitem[{{Faucher-Gigu{\`e}re} {et~al.}(2008){Faucher-Gigu{\`e}re}, {Lidz},
  {Hernquist}, \& {Zaldarriaga}}]{2008FG}
{Faucher-Gigu{\`e}re}, C., {Lidz}, A., {Hernquist}, L., \& {Zaldarriaga}, M.
  2008, \apj, 688, 85

\bibitem[{{Fisher} {et~al.}(2013){Fisher}, {Bolatto}, {Herrera-Camus},
  {Draine}, {Donaldson}, {Walter}, {Sandstrom}, {Leroy}, {Cannon}, \&
  {Gordon}}]{2013Fisher}
{Fisher}, D.~B., {Bolatto}, A.~D., {Herrera-Camus}, R., {Draine}, B.~T.,
  {Donaldson}, J., {Walter}, F., {Sandstrom}, K.~M., {Leroy}, A.~K., {Cannon},
  J., \& {Gordon}, K. 2013, ArXiv e-prints

\bibitem[{{Fontanot} {et~al.}(2014){Fontanot}, {Cristiani}, {Pfrommer},
  {Cupani}, \& {Vanzella}}]{2014Fontanot}
{Fontanot}, F., {Cristiani}, S., {Pfrommer}, C., {Cupani}, G., \& {Vanzella},
  E. 2014, \mnras, 438, 2097

\bibitem[{{Galametz} {et~al.}(2011){Galametz}, {Madden}, {Galliano}, {Hony},
  {Bendo}, \& {Sauvage}}]{2011Galametz}
{Galametz}, M., {Madden}, S.~C., {Galliano}, F., {Hony}, S., {Bendo}, G.~J., \&
  {Sauvage}, M. 2011, \aap, 532, A56

\bibitem[{{Gnedin}(2000)}]{2000aGnedin}
{Gnedin}, N.~Y. 2000, \apj, 535, 530

\bibitem[{{Hahn} \& {Abel}(2011)}]{2011Hahn}
{Hahn}, O., \& {Abel}, T. 2011, \mnras, 415, 2101

\bibitem[{{Inoue} \& {Kamaya}(2008)}]{2008Inoue}
{Inoue}, A.~K., \& {Kamaya}, H. 2008, ArXiv e-prints

\bibitem[{{Iwata} {et~al.}(2009){Iwata}, {Inoue}, {Matsuda}, {Furusawa},
  {Hayashino}, {Kousai}, {Akiyama}, {Yamada}, {Burgarella}, \&
  {Deharveng}}]{iwata09}
{Iwata}, I., {Inoue}, A.~K., {Matsuda}, Y., {Furusawa}, H., {Hayashino}, T.,
  {Kousai}, K., {Akiyama}, M., {Yamada}, T., {Burgarella}, D., \& {Deharveng},
  J.-M. 2009, \apj, 692, 1287

\bibitem[{{Joung} \& {Mac Low}(2006)}]{2006Joung}
{Joung}, M.~K.~R., \& {Mac Low}, M. 2006, \apj, 653, 1266

\bibitem[{{Kimm} \& {Cen}(2014)}]{2014Kimm}
{Kimm}, T., \& {Cen}, R. 2014, \apj, 788, 121

\bibitem[{{Komatsu} {et~al.}(2011){Komatsu}, {Smith}, {Dunkley}, {Bennett},
  {Gold}, {Hinshaw}, {Jarosik}, {Larson}, {Nolta}, {Page}, {Spergel},
  {Halpern}, {Hill}, {Kogut}, {Limon}, {Meyer}, {Odegard}, {Tucker}, {Weiland},
  {Wollack}, \& {Wright}}]{2011Komatsu}
{Komatsu}, E., {Smith}, K.~M., {Dunkley}, J., {Bennett}, C.~L., {Gold}, B.,
  {Hinshaw}, G., {Jarosik}, N., {Larson}, D., {Nolta}, M.~R., {Page}, L.,
  {Spergel}, D.~N., {Halpern}, M., {Hill}, R.~S., {Kogut}, A., {Limon}, M.,
  {Meyer}, S.~S., {Odegard}, N., {Tucker}, G.~S., {Weiland}, J.~L., {Wollack},
  E., \& {Wright}, E.~L. 2011, \apjs, 192, 18

\bibitem[{{Leitherer} {et~al.}(1999){Leitherer}, {Schaerer}, {Goldader},
  {Delgado}, {Robert}, {Kune}, {de Mello}, {Devost}, \&
  {Heckman}}]{1999Leitherer}
{Leitherer}, C., {Schaerer}, D., {Goldader}, J.~D., {Delgado}, R.~M.~G.,
  {Robert}, C., {Kune}, D.~F., {de Mello}, D.~F., {Devost}, D., \& {Heckman},
  T.~M. 1999, \apjs, 123, 3

\bibitem[{{Lisenfeld} \& {Ferrara}(1998)}]{1998Lisenfeld}
{Lisenfeld}, U., \& {Ferrara}, A. 1998, \apj, 496, 145

\bibitem[{{Mostardi} {et~al.}(2013){Mostardi}, {Shapley}, {Nestor}, {Steidel},
  {Reddy}, \& {Trainor}}]{2013Mostardi}
{Mostardi}, R.~E., {Shapley}, A.~E., {Nestor}, D.~B., {Steidel}, C.~C.,
  {Reddy}, N.~A., \& {Trainor}, R.~F. 2013, \apj, 779, 65

\bibitem[{{Nestor} {et~al.}(2013){Nestor}, {Shapley}, {Kornei}, {Steidel}, \&
  {Siana}}]{nestor13}
{Nestor}, D.~B., {Shapley}, A.~E., {Kornei}, K.~A., {Steidel}, C.~C., \&
  {Siana}, B. 2013, \apj, 765, 47

\bibitem[{{Nestor} {et~al.}(2011){Nestor}, {Shapley}, {Steidel}, \&
  {Siana}}]{nestor11}
{Nestor}, D.~B., {Shapley}, A.~E., {Steidel}, C.~C., \& {Siana}, B. 2011, \apj,
  736, 18

\bibitem[{{Osterbrock} \& {Ferland}(2006)}]{2006Osterbrock}
{Osterbrock}, D.~E., \& {Ferland}, G.~J. 2006, {Astrophysics of gaseous nebulae
  and active galactic nuclei}

\bibitem[{{Razoumov} \& {Sommer-Larsen}(2010)}]{2010Razoumov}
{Razoumov}, A.~O., \& {Sommer-Larsen}, J. 2010, \apj, 710, 1239

\bibitem[{{Rosdahl} {et~al.}(2013){Rosdahl}, {Blaizot}, {Aubert}, {Stranex}, \&
  {Teyssier}}]{rosdahl13}
{Rosdahl}, J., {Blaizot}, J., {Aubert}, D., {Stranex}, T., \& {Teyssier}, R.
  2013, ArXiv e-prints

\bibitem[{{Shapley} {et~al.}(2006){Shapley}, {Steidel}, {Pettini},
  {Adelberger}, \& {Erb}}]{shapley06}
{Shapley}, A.~E., {Steidel}, C.~C., {Pettini}, M., {Adelberger}, K.~L., \&
  {Erb}, D.~K. 2006, \apj, 651, 688

\bibitem[{{Teyssier}(2002)}]{teyssier02}
{Teyssier}, R. 2002, \aap, 385, 337

\bibitem[{{Tweed} {et~al.}(2009){Tweed}, {Devriendt}, {Blaizot}, {Colombi}, \&
  {Slyz}}]{2009Tweed}
{Tweed}, D., {Devriendt}, J., {Blaizot}, J., {Colombi}, S., \& {Slyz}, A. 2009,
  \aap, 506, 647

\bibitem[{{Wise} \& {Cen}(2009)}]{2009Wise}
{Wise}, J.~H., \& {Cen}, R. 2009, \apj, 693, 984

\bibitem[{{Yajima} {et~al.}(2014){Yajima}, {Li}, {Zhu}, {Abel}, {Gronwall}, \&
  {Ciardullo}}]{2014Yajima}
{Yajima}, H., {Li}, Y., {Zhu}, Q., {Abel}, T., {Gronwall}, C., \& {Ciardullo},
  R. 2014, \mnras, 440, 776

\end{thebibliography}

\end{document}